\documentclass[conference]{IEEEtran}
\usepackage{graphicx}
\usepackage{placeins}
\usepackage{hyperref}
\usepackage{amsmath}
\usepackage{pifont}
\usepackage{float}
\usepackage{llncsconf}
\usepackage{array}
\usepackage{multirow}

\usepackage{multicol}
\usepackage[per-mode=symbol]{siunitx}
\usepackage{comment}
\usepackage{siunitx}
\usepackage{caption}
\usepackage{subcaption}

\usepackage{graphicx}

\usepackage{siunitx}
\usepackage{url}

\usepackage{cite}
\usepackage{amsmath,amssymb,amsfonts}
\usepackage{algorithmic}
\usepackage{textcomp}
\usepackage{xcolor}

\def\BibTeX{{\rm B\kern-.05em{\sc i\kern-.025em b}\kern-.08em
    T\kern-.1667em\lower.7ex\hbox{E}\kern-.125emX}}

\makeatletter
\newcommand{\newlineauthors}{%
  \end{@IEEEauthorhalign}\hfill\mbox{}\par
  \mbox{}\hfill\begin{@IEEEauthorhalign}
}

\usepackage{tikz}

\newcommand\copyrighttext{%
  \footnotesize \textcopyright \the\year{} IEEE. Personal use of this material is permitted. Permission from IEEE must be obtained for all other uses, including reprinting/republishing this material for advertising or promotional purposes, collecting new collected works for resale or redistribution to servers or lists, or reuse of any copyrighted component of this work in other works.}

\newcommand\copyrightnotice{%
\begin{tikzpicture}[remember picture,overlay]
\node[anchor=south,yshift=10pt] at (current page.south) {\fbox{\parbox{\dimexpr0.75\textwidth-\fboxsep-\fboxrule\relax}{\copyrighttext}}};
\end{tikzpicture}%
}

\def\BibTeX{{\rm B\kern-.05em{\sc i\kern-.025em b}\kern-.08em
    T\kern-.1667em\lower.7ex\hbox{E}\kern-.125emX}}
    
\begin{document}

\title{Efficient Location-Based Service Discovery for IoT and Edge Computing in the 6G Era}
\thanks{Identify applicable funding agency here. If none, delete this.}

\author{\IEEEauthorblockN{Kurt Horvath}
\IEEEauthorblockA{\textit{ITEC} \\
\textit{University of Klagenfurt}\\
Klagenfurt, Austria \\
0009-0008-7737-7013 \\
kurt.horvath@aau.at}
~\\
\and
\IEEEauthorblockN{Dragi Kimovski}
\IEEEauthorblockA{\textit{ITEC} \\
\textit{University of Klagenfurt}\\
Klagenfurt, Austria \\
0000-0001-5933-3246 \\
dragi.kimovski@aau.at}
}

\maketitle   
\copyrightnotice
\begin{abstract}
Efficient service discovery is a cornerstone of the rapidly expanding Internet of Things (IoT) and edge computing ecosystems, where low latency and localized service provisioning are critical. This paper proposes a novel location-based DNS (Domain Name System) method that leverages Location Resource Records (LOC RRs) to enhance service discovery. By embedding geographic data in DNS responses, the system dynamically allocates services to edge nodes based on user proximity, ensuring reduced latency and improved Quality of Service (QoS). Comprehensive evaluations demonstrate minimal computational overhead, with processing times below 1 ms, making the approach highly suitable for latency-sensitive applications. Furthermore, the proposed methodology aligns with emerging 6G standards, which promise sub-millisecond latency and robust connectivity. Future research will focus on real-world deployment, validating the approach in dynamic IoT environments. This work establishes a scalable, efficient, and practical framework for location-aware service discovery, providing a strong foundation for next-generation IoT and edge-computing solutions.
\end{abstract}


\section{Introduction}\label{sub:introduction}

Efficient service discovery plays a pivotal role in enabling communication and resource allocation in the Internet of Things (IoT) and edge computing ecosystems \cite{kimovski2021mobility}. As IoT devices and edge nodes grow rapidly, traditional DNS-based approaches to service discovery encounter limitations in addressing edge nodes to fulfil the demands for low latency, scalability, and adaptability to dynamic user mobility. The increasing complexity of distributed networks in the computing continuum \cite{kimovski21} magnifies these challenges, along with the stringent latency requirements of real-time applications.

The emergence of 6G networks, with their promise of sub-millisecond latency, enhanced reliability, and seamless connectivity, provides a unique opportunity to innovate service discovery methodologies. By embedding geographic information within the DNS infrastructure, we enable location-aware service discovery mechanisms that dynamically allocate services based on user proximity. This capability is critical for edge computing environments, where reducing latency and ensuring localized processing supports next-generation IoT applications such as autonomous vehicles, smart cities, augmented reality, and industrial IoT.

In this work, we propose a novel location-based DNS method that leverages Location Resource Records (LOC RRs) defined by RFC-1034 \cite{rfc1034} to deliver efficient and scalable service discovery. By integrating LOC RRs into DNS responses, we enable the definition of applicable service areas (ASAs) for edge servers, allowing users to connect to the nearest service instance dynamically. This approach improves service localization and minimizes network latency, making it ideal for latency-sensitive IoT and edge computing applications \cite{Pham23}.

Our contributions are threefold:
\begin{itemize}
    \item \textit{Design and Implementation of a Location-Based DNS Framework:}  
    We extend traditional DNS functionality by incorporating LOC RRs to define service areas associated with edge servers. The framework ensures that users connect to the optimal service instance based on proximity, significantly reducing latency and enhancing Quality of Service (QoS).

    \item \textit{Comprehensive Evaluation of Overhead and Performance:}  
    We evaluate the proposed approach, demonstrating that it introduces minimal computational overhead, with processing times consistently below 1 ms (125 LOC RRs). This makes the method highly suitable for real-time, latency-sensitive IoT applications.

    \item \textit{Seamless Alignment with 6G Standards:}  
    We showcase how the proposed method aligns with emerging 6G capabilities, enabling ultra-low latency service discovery and supporting a wide range of dynamic IoT applications within future 6G networks.
\end{itemize}

This paper organizes the content as follows: Section~\ref{sec:related} reviews related work on DNS-based and location-aware service discovery. Section~\ref{sec:methodology} details the methodology, including using LOC RRs and the workflow for service discovery. Section~\ref{sec:evaluation} presents the performance evaluation and results. Finally, Section~\ref{sec:conclusion} provides key insights, concluding remarks, and outlines future directions, including large-scale real-world testing of the proposed system with mobile devices.

\section{Related Work}\label{sec:related}
This section first outlines the methods used for service discovery within our problem domain. It then discusses why DNS serves as an effective mechanism for service discovery, highlighting its inherent advantages. Finally, it explores how location awareness enhances perceived performance, a concept supported by findings from prior methods and studies.

\subsection{Service Discovery for IoT}\label{sub:sdiot}
Service discovery is key in the Internet of Things (IoT), enabling devices to dynamically locate and utilize services in distributed environments. Three main categories of service discovery methods are prevalent: centralized, decentralized, and hybrid. These methods vary in architectures, strengths, and constraints, tailored to address IoT systems' heterogeneous and resource-constrained nature.

\paragraph{Centralized Methods}
Centralized service discovery mechanisms store service information in a central repository or server. An entity queries the server to locate desired services, and the central server responds with appropriate matches. The centralized approach offers fast query processing and simplifies security and access control due to its single control point. However, these mechanisms face significant challenges in terms of scalability and resilience. Jia introduces examples of such methods \cite{jia2017centralized} by introducing a multi-stage semantic service matching algorithm. Hamrouni \ proposes another method \cite{hamrouni2022context}, introducing a context-aware service discovery method using graph theory and socially connected objects.

The reliance on a central data source creates a single point of failure, which can compromise the entire system if the central server becomes unavailable. Furthermore, centralized systems struggle with high query traffic, unreliable latency and bottleneck issues as IoT deployments scale. Heidari \cite{heidari2022service} noted in his work that while centralized approaches are efficient for small-scale networks, they are unsuitable for many IoT systems' dynamic and large-scale nature.

\paragraph{Decentralized Methods}
Decentralized mechanisms eliminate the reliance on a central server by distributing the service discovery process across multiple nodes in the network using multiple approaches to how they relate to each other. In those approaches, devices collaborate to locate services, which enhances scalability and fault tolerance. Peer-to-peer (P2P) systems and agent-based frameworks are prominent examples of decentralized service discovery. P2P systems, such as those employing semantic overlays, improve search efficiency and reduce reliance on central nodes, enabling better scalability \cite{parhi2018multi}. Similarly, agent-based mechanisms leverage autonomous agents to dynamically negotiate and coordinate service discovery. Another category is broker-concept-based systems, as introduced by Nguyen \cite{nguyen2022bmdd}, who introduces BMDD. Brokers per se are centralized components, but methods like BMDD use the same architecture and create the same workflow on a federated level. Another method proposed by Murturi \cite{murturi2021decentralized} places metadata information on edge nodes and enables clients to discover other edge nodes. Many service discovery schemes focus on device-to-device interaction or device-to-nodes. Still, Tang\cite{tang2021iota}, for example, dives deeper into the computing continuum towards the fog domain to provide higher computational capabilities.

These decentralized methods excel in dynamic IoT environments but often incur higher complexity and security challenges than centralized systems.

\paragraph{Hybrid Methods}
Hybrid mechanisms combine the strengths of centralized and decentralized approaches to balance efficiency and scalability. By integrating centralized control for critical tasks and decentralized collaboration for local service discovery, hybrid systems address some of the limitations inherent in the other two categories. For instance, Cheng \cite{cheng20} demonstrated that hybrid architectures reduce network traffic by localizing service changes while ensuring global service availability. Also notable is the work of Tsai \cite{tsai2011hybrid}, who also used a hybrid method not for the edge but for web services in general. However, these systems can be complex and require careful coordination between centralized and decentralized components.

\subsection{DNS for Service Discovery}\label{sub:dns4sd}
DNS-based service discovery in IoT environments relies on lightweight and efficient protocols such as Multicast DNS (mDNS) \cite{rfc6762} and DNS Service Discovery (DNS-SD) \cite{rfc6763}. These solutions enable IoT devices to find services locally and globally within networks. Multicast DNS (mDNS) facilitates the resolution of hostnames to IP addresses without the need for a conventional DNS server, which makes it particularly suitable for local networks in IoT settings. It achieves this by broadcasting queries to all devices on the same subnet, thus supporting decentralized and lightweight discovery. However, mDNS introduces high multicast traffic, which can cause significant network congestion, especially in large-scale IoT deployments.
DNS Service Discovery (DNS-SD) extends the capabilities of standard DNS by enabling devices to advertise their services using metadata-rich service records. This approach leverages the DNS infrastructure's hierarchical and scalable nature to allow IoT devices to discover services across networks efficiently. DNS-SD excels in its interoperability and scalability, integrating seamlessly with existing DNS infrastructure.
Various Methods like \cite{kaiser14} introduced by Kaiser and \cite{siljanovski2014service} by Siljanovski directly build up on DNS-SD to adapt service discovery for their particular domains.

Others like Zang \cite{zhang2017novel} and Djamaa \cite{Djamaa14} have highlighted the utility of DNS-based solutions in meeting IoT scalability and interoperability needs while noting the critical gaps in security and energy efficiency.
Other methods more tailored for the computing continuum and applicable to IoT applications are the location-aware service discovery methods proposed by Horvath et al. \cite{mesddhorvath23} \cite{horvath2022location}.
These observations emphasize the necessity for innovation to address the specific constraints of IoT environments, as many IoT applications depend on low latency. The service discovery method and access networks, such as 5G and 6G, must also support this requirement \cite{ma2019high}.

\subsection{Location-based Services Discovery}\label{sub:locationbasedServices}
Incorporating user location information into service discovery algorithms can also benefit service providers by allowing them to manage their resources better. Providers can better distribute the load across their servers by directing requests to the closest edge node, resulting in more efficient resource utilization \cite{cheng2019smart}. Furthermore, location-based service discovery can help providers identify areas where their services may be in high demand, allowing them to strategically deploy resources to meet those demands \cite{cheng2019smart}.

However, challenges are also associated with incorporating user location information into service discovery. One major challenge is the issue of user privacy, as collecting and using location data can raise concerns about the misuse of personal information \cite{wang2019location}. Service providers must implement appropriate measures to protect user privacy, such as anonymizing location data and obtaining explicit user consent to use their location information.

In addition, incorporating location information into DNS can require changes to the existing infrastructure and protocols, which may involve significant coordination and standardization efforts \cite{huang2018toward}. Furthermore, geolocation data may not always be accurate or available, particularly for users accessing the internet through a virtual private network (VPN) or other network configurations \cite{belik2017ip}. Service providers must consider these limitations and develop strategies to handle cases where location data may be unreliable or unavailable.

Despite these challenges, incorporating user location into service discovery holds great potential for improving the efficiency and effectiveness of Internet services. Continued research and development in this area is essential to harness the benefits of location-based service discovery fully.

\subsection{LOC Ressource Records for DNS}\label{sub:LOCrr}

The Geographical Location (LOC) Resource Record (RR) type in DNS can represent locations. The LOC RR allows geographic information, such as latitude and longitude, to be associated with a DNS domain name.

The LOC RR is defined in \textit{RFC 1876} \cite{rfc1876} and is structured as follows:

\begin{verbatim}
<owner> <ttl> IN LOC <latitude> 
<longitude> [<altitude>]
\end{verbatim}

\begin{itemize}
    \item \texttt{<owner>} is the domain name associated with the LOC RR.
    \item \texttt{<ttl>} is the time to live for the record.
    \item \texttt{<latitude>} and \texttt{<longitude>} are the coordinates of the location. These are specified in decimal degrees using the WGS 84 \cite{gomarasca2009basics} format.
    \item \texttt{<altitude>} (optional) is the altitude of the location in meters above sea level.
\end{itemize}
For better understanding, we provide an example:
\begin{verbatim}
edgeA.myservice.com LOC 28 43 0.0 N 
128 12 0.0 W 200.00m 20m 10m 10m
\end{verbatim}
\begin{itemize}
    \item \textit{Host:} \texttt{edgeA.myservice.com}
    \item \textit{Latitude:} $28^\circ 43' 0.0'' \mathrm{N}$ (28 degrees, 43 minutes North)
    \item \textit{Longitude:} $128^\circ 12' 0.0'' \mathrm{W}$ (128 degrees, 12 minutes West)
    \item \textit{Altitude:} $200.00 \, \mathrm{m}$ (200 meters below sea level)
    \item \textit{Size:} $20 \, \mathrm{m}$ (sphere diameter of 20 meters)
    \item \textit{Horizontal Precision:} $10 \, \mathrm{m}$ (latitude/longitude accuracy within 10 meters)
    \item \textit{Vertical Precision:} $10 \, \mathrm{m}$ (altitude accuracy within 10 meters)
\end{itemize}

DNS-based location representation can complemented using other types of DNS records, such as the Address (A) and Pointer (PTR) or Canonical Name (CNAME) records defined in RFC-1034 \cite{rfc1034}. For example, a CNAME can be linked with a LOC to represent a location.

\section{Methodology}\label{sec:methodology}
First, we formalize how applicable service areas (ASAs) relate to service instances, which will be outlined in the Model. The Model also specifies the algorithm used to verify whether a user resides within an ASA. Following this, we present the general workflow of our service discovery (SD) method. Finally, we provide topological definitions to describe not only the correlation between servers and service areas but also how to define a layout effectively.

\subsection{Model}\label{sub:model}
We define a service $S$ as a set of service instances that provision either cloud-hosted or edge-hosted instances: $S = (S^{cloud}_i, S^{edge}_j)$, where $i$ and $j$ represent the distinct hosts of each type. Each edge service instance manages a set of ASAs - $A$, where such area is defined as follows:

\begin{equation} \label{eq:area}
\begin{aligned}
A_k = (P, r, S^{edge}_j)\\
R \subseteq A_k \times S^{edge}_j
\end{aligned}
\end{equation}

Here, $P$ represents a point in Cartesian coordinates defined by latitude and longitude, $r$ specifies the radius that defines the circular expanse of the area, and $S^{edge}_j$ denotes the associated edge instance.

We further specify that each $S^{edge}_j$ can be assigned to multiple ASAs, as expressed below:

\begin{equation} \label{eq:relation}
\begin{aligned}
\forall S_j, \exists A_k \text{ (with \(k \neq k_{+1}\) allowed) such that } \\(A_k, S_j), (A_{k+1}, S_j) \in R.
\end{aligned}
\end{equation}

This assignment respects the constraint that each service instance must be associated with at least one area, defined as follows:

\begin{equation} \label{eq:constraints}
\begin{aligned}
\forall S_j, \exists A_k \text{ such that } (k, j) \in R.
\end{aligned}
\end{equation}

When a user $u$ is near an area $A_k$, the corresponding edge service instance from the relation $R$ must provide the service. We define this relationship as follows:

\begin{equation} \label{eq:relationedgeuser}
\begin{aligned}
\forall u \in A_k, \exists S^{edge}_j: \begin{cases} 
\texttt{true} & u \rightarrow S^{edge}_j, \\
\texttt{false} & u \rightarrow S^{cloud}_i 
\end{cases} \forall S^{edge}_j, S^{cloud}_i \in R,
\end{aligned}
\end{equation}

If the user is not within any area $A^{edge}$, one of the cloud service instances $S^{cloud}_i$ must serve the user.

\paragraph{Locate user in areas}\label{sub:userinArea}
Identifying the areas where a user resides can be a more computationally complex task. This process requires two steps. First, transform the spherical coordinates used for the location into cartesian coordinates. According to the WGS 84 format\cite{gomarasca2009basics} we define as follows:
\begin{itemize}
    \item Latitude: \(\phi\) (in degrees, \(-90^\circ \leq \phi \leq 90^\circ\))
    \item Longitude: \(\lambda\) (in degrees, \(-180^\circ \leq \lambda \leq 180^\circ\))
    \item Ellipsoidal height: \(h\) (in meters), derived from the user's elevation.
\end{itemize}

Subsequently we compute the Cartesian coordinates \((x, y)\) as follows \cite{grewal2020global}:
\begin{equation} \label{eq:xy}
\begin{aligned}
x = (N + h) \cos(\phi) \cos(\lambda),\\
y = (N + h) \cos(\phi) \sin(\lambda).
\end{aligned}
\end{equation}

Here, $N$, the radius of curvature in the prime vertical, is defined as:
\begin{equation} \label{eq:radius}
\begin{aligned}
N = \frac{a}{\sqrt{1 - e^2 \sin(\phi)^2}},
\end{aligned}
\end{equation}
where $a = 6378137$ meters and $e = 0.01671$ the eccentricity of the Earth's ellipsoid.
Next, in the second step, we determine whether the user is inside or outside the defined area:
\begin{equation} \label{eq:metricINout}
\begin{aligned}
\sqrt{(x_u - x_{loc})^2 + (y_u - y_{loc})^2} \leq r = \begin{cases} 
\texttt{true} & \text{INSIDE}, \\
\texttt{false} & \text{OUTSIDE} \\
\end{cases}
\end{aligned}
\end{equation}

In this equation, \(x_u, y_u\) represents the user's position, and \(x_{loc}, y_{loc}\) represents the position specified in the LOC resource record.

\subsection{Workflow}\label{sub:workflow}
One of the main design goals when using LOC RR to locate a service instance is to minimize the overhead compared to conventional service discovery, which, in its simplest form, involves a DNS lookup. By definition, DNS is a strict hierarchical name resolution method that propagates resolvable addresses and servers within this hierarchy to end users who want to consume a service under a given name. A typical request requires only a Uniform Resource Locator (URL) as specified in RFC-1034 \cite{rfc1034}, and the response provides either \textit{CNAME records} as aliases or \textit{A records} with the specific service address to access the service.
\begin{figure}[t]
     \centering
     \includegraphics[width=0.5\textwidth]{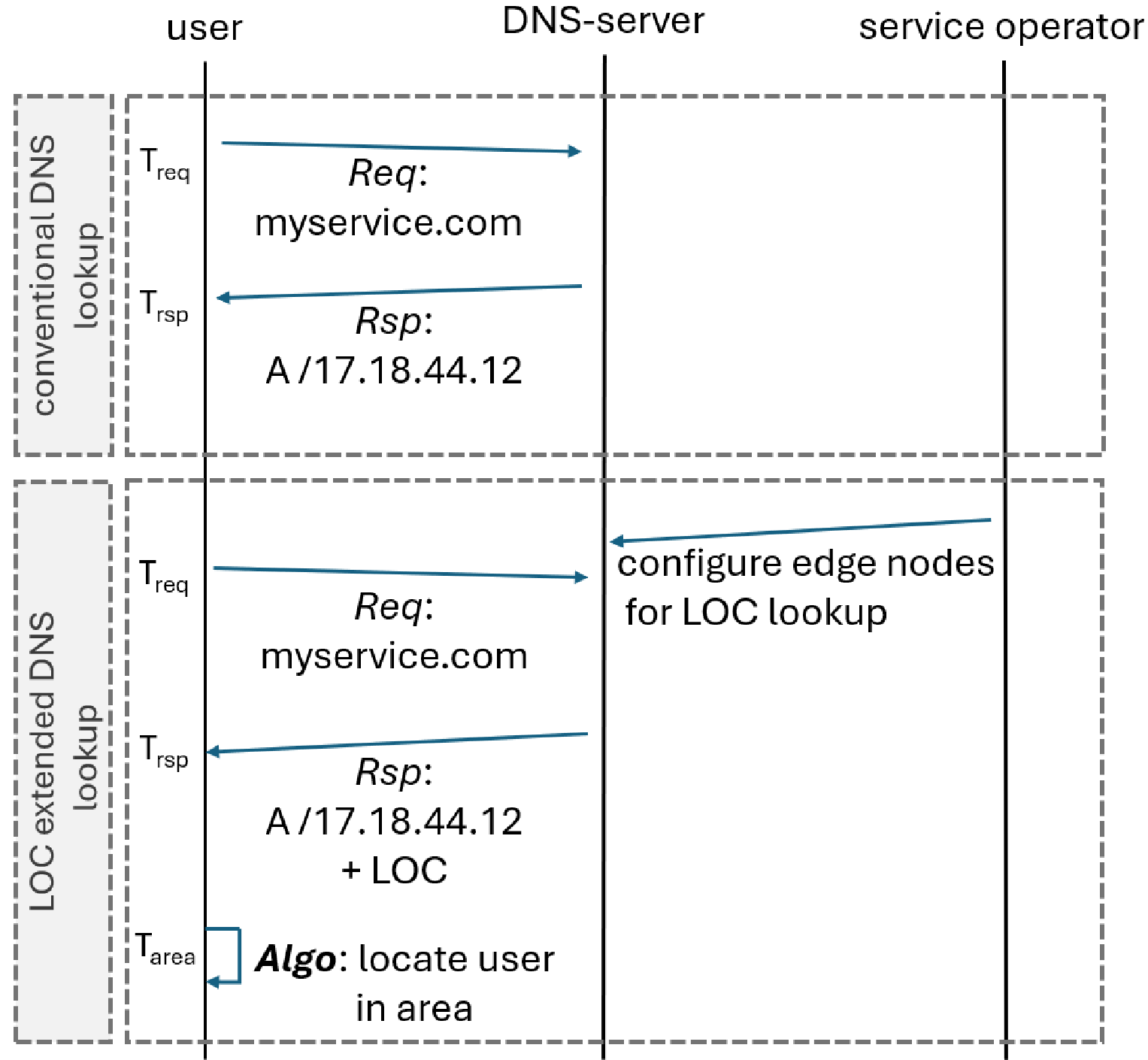} %
     \caption{Workflow of service discovery}%
     \label{fig:workflow}%
\end{figure}
Figure \ref{fig:workflow} illustrates the conventional DNS lookup process. We enhance this process by incorporating \textit{LOC records}, which designate specific geographic areas (ASAs) for particular service instances. To implement this, the service operator must first configure the available areas and assign corresponding service instances, as defined in Section \ref{sub:model}.
After establishing the configuration, DNS requests from users return not only the conventional \textit{A records} and \textit{CNAME records} but also a list of \textit{LOC records} as described in Section \ref{sub:LOCrr}, that indicate the locations where an IP address is applicable. The client receives this comprehensive list of Applicable Service Areas (ASAs) and evaluates its location relative to the defined areas. This evaluation introduces additional computational complexity but does not add more requests as overhead. The added time, denoted as $T_{area}$, stems from the increased payload size in the DNS request/response exchange, while $T_{req}$ and $T_{rsp}$ remain unaffected, as depicted in Figure \ref{fig:workflow}.
After receiving the DNS response, the client processes the payload by first extracting the \textit{CNAME records} and \textit{A records}, followed by the \textit{LOC records}. The client then applies the algorithm described in Section \ref{sub:userinArea} to determine whether the user resides within specified areas. This streamlined workflow ensures efficient location-aware service discovery while maintaining compatibility with existing DNS infrastructure.

\subsection{Topology}\label{sub:topology}
\begin{figure}[t] 
    \centering 
    \includegraphics[width=0.5\textwidth]{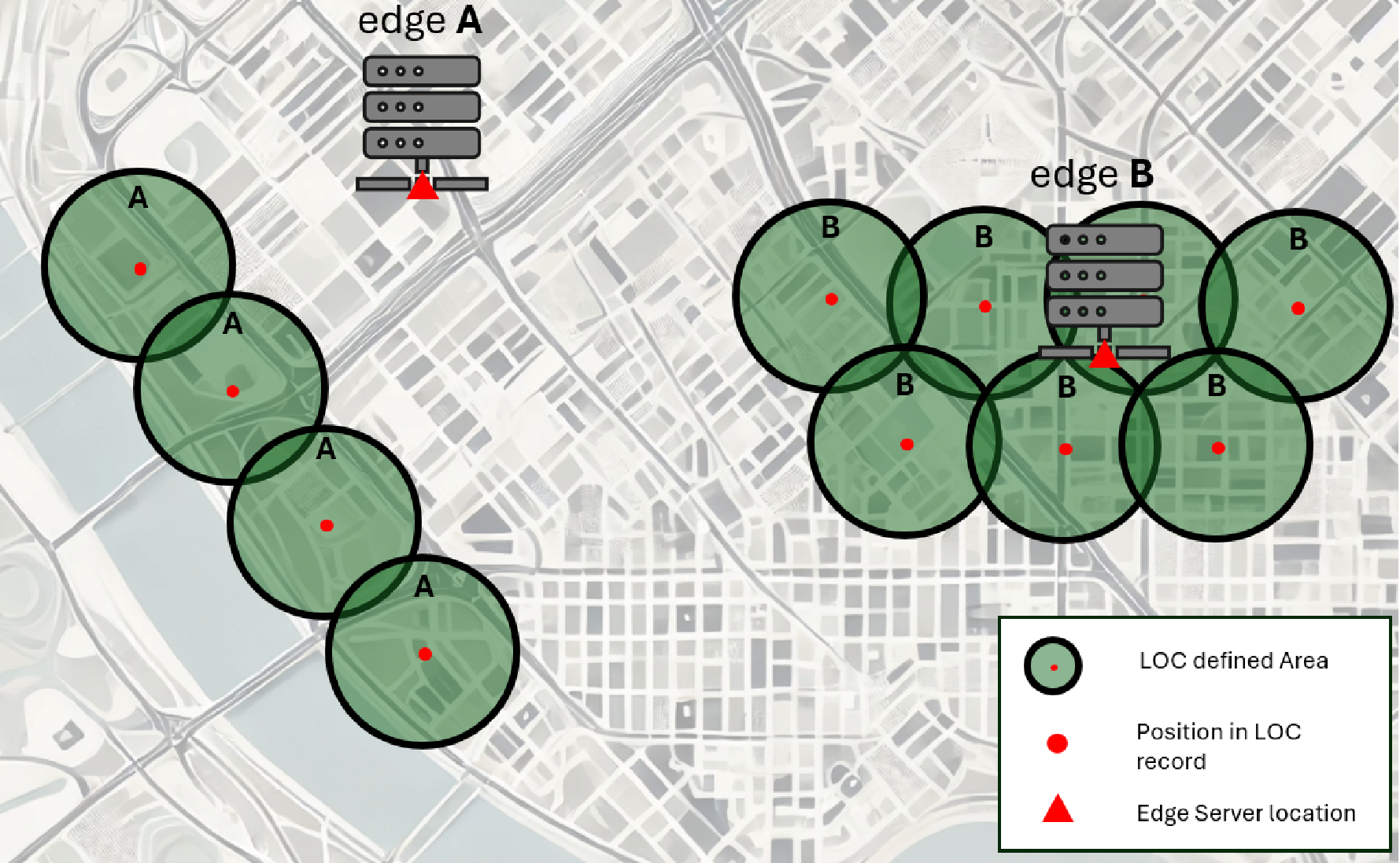} 
    \caption{The alignment of applicable service areas to their edge nodes is organized in a specific topology.}
    \label{fig:concept} 
    \end{figure}

The topology determines how to define areas that align with edge server instances. Figure \ref{fig:concept} shows two edge servers (\texttt{A} and \texttt{B}) and eleven areas in total. Edge A handles the four areas on the left, while \texttt{edge B} manages the remaining seven. A radius of 500 meters defines the size of the area.
As Figure \ref{fig:concept} illustrates, a service instance node does not need to reside inside the area it serves. Instead, placement depends on connectivity, resource availability, and demand. For instance, if an area, such as a coastline, requires low latency due to specific use cases,  the service operator can define areas accordingly and select a server that ensures satisfactory QoS. 

The algorithm described in Equation \ref{eq:relationedgeuser} assigns users to the correct area by applying a search process for each area. Overlapping areas are allowed because the search stops once it identifies the first area the user occupies. This overlap ensures greater flexibility, especially in dynamic environments where user locations and demands frequently shift. For example, when a user moves from one area to another, the algorithm quickly identifies the new area (Equation \ref{eq:radius}). It reassigns the user to the appropriate edge server without noticeable delay.

Moreover, system operators can configure the topology to prioritize certain areas based on business or operational needs. For instance, areas with higher population densities or critical infrastructure might receive more edge server resources to guarantee seamless service. Operators can enhance this adaptability by assigning temporal validity to the defined areas, a concept already in DNS.

\section{Evaluation}\label{sec:evaluation}
This section explains the selected metrics for evaluation, their relevance, and the results. The evaluation aims to assess our proposed method's efficiency, accuracy, and scalability under different conditions.

\subsection{Metrics}\label{sub:metrics}
As outlined in Section \ref{sub:workflow}, our method builds upon a conventional DNS lookup workflow. This workflow involves two key steps influencing the overall performance and user experience. Evaluating these steps allows us to understand the feasibility and practical implications of the method.

First, the client obtains its location through a GPS position. Acquiring the current position in precise detail can take between 100 and 500 ms, depending on the hardware and environmental conditions, such as GPS signal availability. However, most operating systems provide optimized solutions for this process. For example, Microsoft's \textit{Location Service} or the Android API allows retrieval of the last known position in less than 1 ms. These APIs are particularly useful in scenarios where real-time precision is not critical. Assuming an area with a diameter of 1000 meters and a maximum user velocity of 50 km/h in urban areas, the inaccuracy introduced by relying on the last known position remains negligible. This optimization ensures a faster response time, critical for applications requiring low latency.

The second step involves determining the user's current area, which requires evaluating the relationship between the user's position and predefined areas. As the number of configured areas increases, the computational effort needed to verify all possible areas grows proportionally. This step becomes a critical performance bottleneck when the system handles many areas. To address this, we define an efficient algorithm, outlined in Section \ref{sub:userinArea}, that reduces processing overhead while maintaining accuracy. This algorithm evaluates whether the user resides in a specific area and forms the cornerstone of our method's computational improvements.

To quantify the performance of this algorithm, we focus on its processing time, designated as $T_{area}$, also depicted in Figure \ref{fig:workflow}. This metric captures the efficiency of the area-determination step and serves as an essential indicator for scalability. We evaluate $T_{area}$ under varying conditions, such as different numbers of areas (LOC list length) and user movement patterns, to identify its behaviour in both optimal and challenging scenarios. Additionally, we measure its impact on the overall system performance, particularly in environments with dynamic user positions and high query loads.

Beyond performance, we also consider the system's robustness and adaptability. For instance, urban environments often introduce challenges such as signal interference and high user density, affecting GPS accuracy and area computation. Our method ensures reliable operation under suboptimal conditions by incorporating fallback mechanisms, such as using approximate locations or caching previously determined areas. These enhancements further optimize user experience by reducing delays and maintaining consistent performance.

We evaluate the end-to-end performance using various public DNS servers to provide a clear perspective on the overhead introduced by the service discovery process. This evaluation delivers concrete values that quantify the "cost" of applying service discovery. As shown in Figure \ref{fig:workflow}, we define the DNS query time $T_q$ as:

\begin{equation} \label{eq:totaltime}
\begin{aligned}
T_q = T_{req} + T_{rsp} + T_{area}.
\end{aligned}
\end{equation}

\subsection{Results}\label{sub:results}
\begin{figure}[t]
     \centering
     \includegraphics[width=0.5\textwidth]{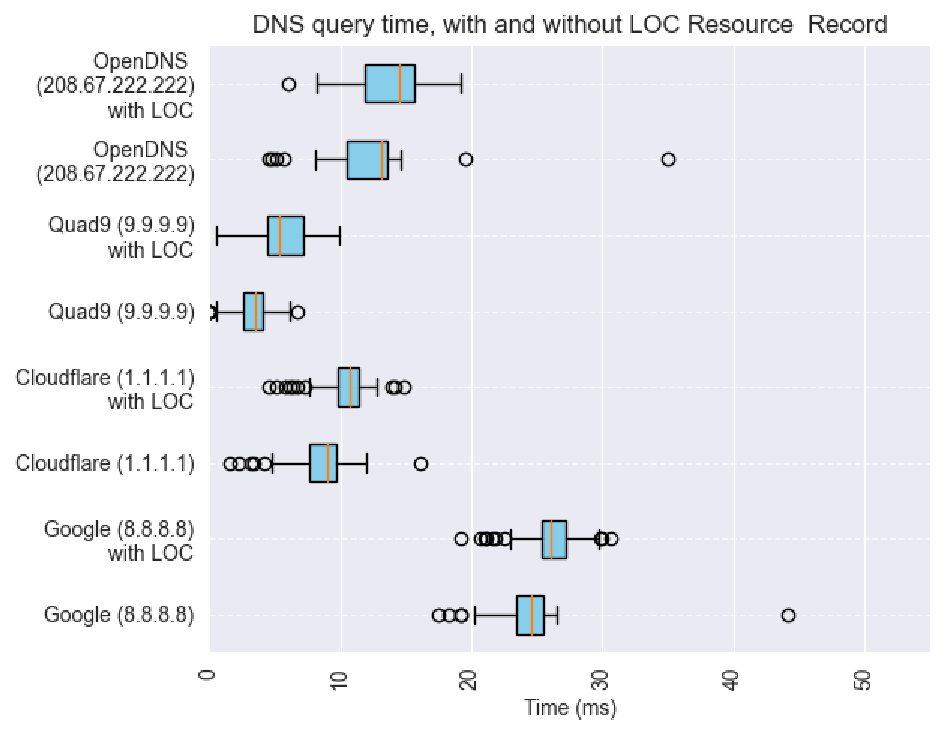} %
     \caption{DNS lookup with and without LOC data.}%
     \label{fig:tboxLOC}%
\end{figure}

\paragraph{End to End Performance}\label{sub:end2end}
To evaluate the impact of appending the LOC Resource record to DNS responses. We selected the four most common public DNS servers and compared their performance. We address the DNS query time, which includes the DNS Request $T_{req}$ and DNS Response $T_{rsp}$. The list of areas provided as LOC records is validated using the Algorithm proposed in Section \ref{sub:userinArea}. Figure \ref{fig:tboxLOC} shows pairs of DNS servers where we conducted evaluations using 100 requests on each server with a 3-second delay.

\textbf{OpenDNS} experiences a 1.6 ms rise in average query time when $T_q$ including LOC records (without LOC Mean: 12.38 ms; with LOC Mean: 13.98 ms). The standard deviation increases slightly, moving from 3.10 ms to 4.35 ms. Interestingly, the absence of LOC records produces the highest outlier, measured at 35.02 ms.

\textbf{CloudFlare} maintains consistent resolution performance, showing only a 0.1 ms increase in the mean time when LOC records are included (without LOC Mean: 13.18 ms; with LOC Mean: 13.28 ms). The standard deviation changes minimally, growing from 2.97 ms to 3.06 ms.

\textbf{Google Public DNS} records a more noticeable impact, with the mean resolution time increasing by 2.7 ms when using LOC records (without LOC Mean: 10.60 ms; with LOC Mean: 13.30 ms). The standard deviation also rises significantly, from 1.72 ms to 4.06 ms.

\textbf{Quad9} registers the most significant increase in mean resolution time with LOC records, reporting a 3.6 ms difference (without LOC Mean: 11.42 ms; with LOC Mean: 15.02 ms). Its standard deviation also sees a marked rise, increasing from 2.04 ms to 5.40 ms.

This evaluation highlights that while LOC records enable location-aware service selection, their integration often incurs performance costs (even small ones), particularly in higher latency and more significant variability in resolution times. These impacts vary significantly across DNS providers.

\begin{figure}[t]
     \centering
     \includegraphics[width=0.5\textwidth]{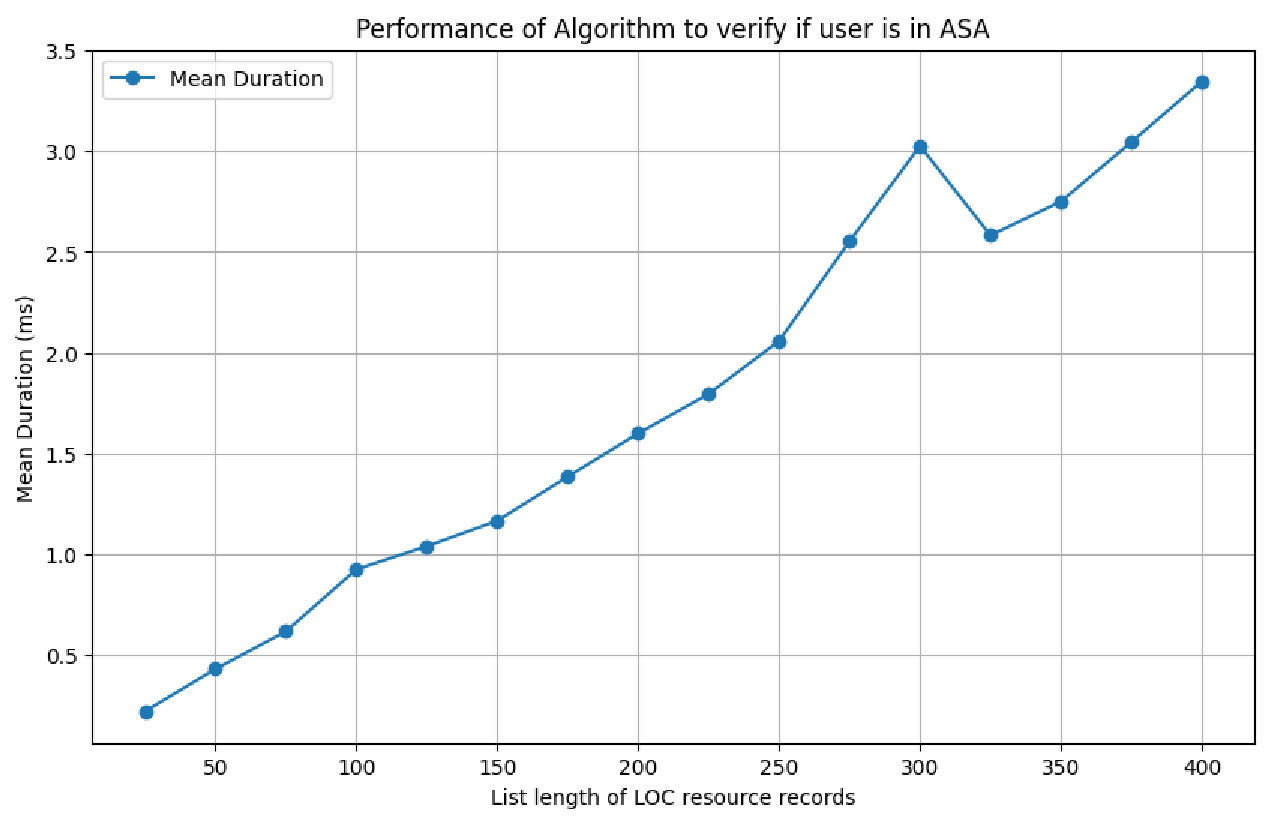} %
     \caption{Performance to identify the area the user resides in.}%
     \label{fig:trace}%
\end{figure}

\paragraph{Locate user in Areas}\label{par:locateuserarea}
Figure \ref{fig:trace} illustrates our evaluation of whether a user resides in an area using a location resource records (LOC-RR) dataset of 400 locations representing approximately 20 km². We implemented the algorithm in Python and conducted experiments to measure the performance on a 11th Gen Intel(R) Core(TM) i7-11800H @ 2.30GHz with 32 GB of RAM and MS Windows 11. 

The results show average processing times for increments of 25 areas, with 100 iterations performed for each LOC list length. This process yielded a total of 1600 measurements, enabling a comprehensive assessment. The results reveal a relatively linear growth in perceived latency, with minor increases observed at 150 elements in an LOC list and noticeable improvements at 300 elements per list.

Table \ref{table:performance_metrics} presents detailed performance metrics for various LOC resource records list lengths. The mean processing time starts at 0.219 ms for 25 areas and increases to 3.347 ms for 400 areas, with maximum durations reaching 9.42 ms. These results highlight that our approach consistently operates below 1 ms as long as the list length of LOC resource records remains under 125 elements. The standard deviation stays minimal, except for an anomaly at a list length of 275. Standard deviation varies from 0.063 ms to 2.46 ms across different configurations, underscoring the stability and predictability of our method. This stability proves crucial in dynamic environments, ensuring consistent performance despite fluctuations in the number of areas or user location changes.
The results indicate linear growth in calculation time until the evaluation of 275 and 300 LOC resource records. At this point, the processing time slightly increases, stabilizing at 325 records. The algorithm maintains its efficiency even with list lengths exceeding 350 elements. This anomaly likely results from memory optimization when the aggregated data in the ROC record list exceeds a cache level.

\begin{table}[]
\resizebox{\columnwidth}{!}{%
\begin{tabular}{|c|c|c|c|c|c|}
\hline
LOC list Length & Mean (ms) & Median (ms) & Max (ms) & Std Dev (ms) & 90th Perc. (ms) \\ \hline
25          & 0.219190   & 0.210125   & 0.784875  & 0.063657   & 0.213609              \\ \hline
50          & 0.429369   & 0.409503   & 1.189540  & 0.088116   & 0.456634              \\ \hline
75          & 0.615968   & 0.604352   & 0.811360  & 0.033449   & 0.665332              \\ \hline
100         & 0.924043   & 0.791955   & 13.691530 & 1.288401   & 0.871947              \\ \hline
125         & 1.039020   & 0.950447   & 6.290520  & 0.557340   & 1.032670              \\ \hline
150         & 1.165155   & 1.129685   & 2.294545  & 0.130105   & 1.202807              \\ \hline
175         & 1.384092   & 1.338570   & 2.267105  & 0.137477   & 1.449013              \\ \hline
200         & 1.598960   & 1.563078   & 2.499760  & 0.121737   & 1.668214              \\ \hline
225         & 1.794737   & 1.770247   & 2.617965  & 0.140413   & 1.883193              \\ \hline
250         & 2.057804   & 2.017735   & 2.829855  & 0.160292   & 2.189199              \\ \hline
275         & 2.552673   & 2.170783   & 18.782710 & 2.299011   & 2.311002              \\ \hline
300         & 3.026073   & 2.370285   & 19.500810 & 2.465071   & 2.959922              \\ \hline
325         & 2.582315   & 2.537343   & 4.895540  & 0.274821   & 2.661656              \\ \hline
350         & 2.750254   & 2.724578   & 3.367070  & 0.116644   & 2.830519              \\ \hline
375         & 3.045138   & 3.008620   & 4.339275  & 0.214894   & 3.243129              \\ \hline
400         & 3.346753   & 3.249782   & 9.426970  & 0.635999   & 3.536691              \\ \hline
\end{tabular}
}
\caption{Performance metrics for different list lengths.}
\label{table:performance_metrics}
\end{table}

Beyond the numerical results, our evaluation underscores the system's practicality and reliability in real-world scenarios. Even under the maximum tested configuration of 400 areas, the processing time remains below 3.53 ms at the 90th percentile, ensuring responsiveness under high-load conditions. The low standard deviation further validates the system's ability to produce stable and predictable outcomes, critical for maintaining a seamless user experience.
This predictability becomes particularly valuable in environments where timing precision directly impacts user satisfaction, such as real-time applications, dynamic resource allocation, or location-based services. By maintaining minimal variability, the system effectively reduces the risk of performance bottlenecks or erratic behaviour, even as the complexity of the task scales. These findings demonstrate the robustness of the approach and its suitability for deployment in dynamic, high-demand scenarios.

\section{Conclusion}\label{sec:conclusion}
Our work illustrates that including location information within DNS responses introduces minimal computational overhead, with processing times consistently measured below 1 ms. This low-latency performance demonstrates that the proposed method integrates seamlessly into existing DNS workflows while providing enhanced functionality. The approach effectively facilitates location-based service discovery by leveraging GEO Resource Records (LOC RRs) to associate geographic locations with specific services. This capability allows users to seamlessly connect to the most optimal service instance, ensuring reduced latency and improved user experience, particularly in distributed and edge-based environments.

The approach is particularly well-suited for edge computing, where efficient and localized service discovery is a cornerstone for supporting dynamic, latency-sensitive applications. The method offers flexibility and scalability by allowing service operators to define applicable service areas and assign edge servers accordingly. Its robustness in handling overlapping areas and user mobility further enhances this, ensuring continuous service delivery without noticeable delays.

The proposed method becomes even more relevant as the next-generation communication standard 6G emerges with its promise of ultra-low latency (sub-1 ms) and enhanced connectivity. 6G technology is expected to revolutionize edge computing by enabling real-time applications such as autonomous vehicles, augmented reality, and industrial IoT \cite{kitanovoverview2024}. The minimal latency overhead of this location-based DNS method makes it an ideal enabler for such 6G-driven use cases, where even slight delays can significantly impact performance.
Further exploration is required to evaluate its full potential in real-world environments. Future work will focus on deploying the complete workflow in practical settings, including extensive testing with real mobile devices under varying conditions. This will involve assessing the system's performance in diverse urban and rural scenarios, examining its adaptability to different hardware and network configurations, and ensuring its resilience under high user loads.

\subsection{6G enabling new services}\label{sub:6gnewservices}
The emergence of 6G networks will be pivotal in supporting next-generation IoT services, which demand ultra-low latency, real-time responsiveness, and dynamic adaptability. IoT applications such as 
 augmented reality, smart cities, and industrial automation require communication networks capable of processing massive data volumes while ensuring < 1 ms latencies \cite{salh2021survey}. These requirements go beyond the capabilities of existing network technologies, positioning 6G as a foundational enabler for such advanced services \cite{kitanovoverview2024}.

Achieving these low latency goals necessitates advanced communication infrastructure, efficient service discovery mechanisms, and edge computing capabilities. Service discovery systems must dynamically connect users to nearby edge servers, ensuring localized data processing and minimizing transmission delays. Integrating geographic information, such as the method we introduce, enhances this by enabling rapid identification and allocation of the optimal edge node.
Edge computing complements this effort by offloading computational tasks from centralized cloud systems to distributed edge nodes closer to the users. This proximity significantly reduces the computational latency component in end-to-end workflows. 

Since the introduction of 5G, researchers have proposed integrating low-latency services through Mobile Edge Computing (MEC), as discussed by Hammer \cite{hammer2023scalable}. Building on bringing computing resources closer to users, User Plane Functions (UPFs) \cite{jain2022evolving} places services between the radio network and the publicly accessible internet. Researchers are actively exploring how to implement service discovery in this area conveniently and effectively.

6G's ability to combine ultra-reliable low-latency communication (URLLC) \cite{siddiqui2023urllc} with scalable edge computing and efficient service discovery forms the backbone for future IoT services. Together, these technologies ensure that user demands for responsiveness and service quality are met, even in dynamic and resource-constrained environments.

\subsection{Future work}\label{sub:futurework}
Our work and emerging technology inspire new research opportunities, particularly in integrating advanced security measures to protect user privacy in location-based DNS services. Researchers must address data anonymity and ensure compliance with global data protection standards to drive widespread adoption of this technology.
As the 6G standard evolves and stakeholders make foundational technology decisions, they must define specific requirements for the communication medium and supplementary services. This includes addressing data persistence, security, service discovery, and user matching \cite{sabella2016mobile}. Current developments indicate that not all service participants will reside within the same network. Mobile providers often impose vendor lock-in, hosting ultra-low-latency applications within their networks \cite{zhang2021internet}.
We must design methods that enable user interactions within locked networks while making those interactions seamless across network operator boundaries. One emerging concept involves User Plane Functions (UPFs), which provide a direct gateway to the User Equipment (UE) to achieve latencies below 1 ms. These UPFs also include a secondary gateway, allowing external users from the public internet to access the same edge service with high, but still moderate, latency.
Service discovery methods must achieve low latency and facilitate communication between public and closed networks in order to design such services. This challenge highlights a critical research gap that requires attention.
\newpage

\section*{Acknowledgement}
This work received funding the Austrian Research Promotion Agency (FFG) (grant 888098, K{\"a}rntner Fog). Supported by the OeAD for collaboration of the University of Klagenfurt (Austria) and Mother Teresa University (Republic of North Macedonia)

\bibliographystyle{IEEEtran}
\bibliography{references}
\end{document}